\documentclass[aps,pra,showpacs,amssymb,nofootinbib,twocolumn]{revtex4}
\usepackage{graphicx}
\begin{document}

\title{Dealing with entanglement of continuous variables:\\Schmidt decomposition with discrete sets of orthogonal functions}
\author{Lucas Lamata}\email{lamata@imaff.cfmac.csic.es}\affiliation{Instituto de
Matem\'aticas y F\'{\i}sica Fundamental, CSIC\\Serrano 113-bis,
28006 MADRID, Spain}
\author{Juan Le\'on}\email{leon@imaff.cfmac.csic.es}
 \affiliation{Instituto de
Matem\'aticas y F\'{\i}sica Fundamental, CSIC\\Serrano 113-bis,
28006 MADRID, Spain}

\begin{abstract}
We propose a method for obtaining the Schmidt decomposition of
bipartite systems with continuous variables. It approximates the
modes to the prescribed accuracy by well known orthogonal functions.
We give some criteria for the control of errors. We illustrate the
method comparing its results with the already published analysis for
entanglement of biphotons. The agreement is excellent.
\end{abstract}
 \maketitle
\section{Introduction\label{s1}}
Bipartite and multipartite entanglement is one of the features
that give rise to many of the developments of quantum computation
and information, like quantum teleportation \cite{tele1,tele2} and
quantum cryptography \cite{crypto1,crypto2}, among others
\cite{nielsen}. The evaluation of the entanglement of a composite
state is thus a main task to be done. For this purpose, the
Schmidt decomposition \cite{schmidtdisc1,schmidtdisc2} has proven
to be a valuable tool, for systems with just two components.

In this paper we consider the case of continuous variables
entanglement. For us, these variables may be
$\{a+a^{\dag},i(a^{\dag}-a)\}$ which commute as phase space
variables do. We also refer to continuous variable entanglement in
systems described by momentum and/or energy observables. Precisely,
the entanglement of continuous variables stems from the original EPR
article \cite{epr}. However, the treatment of systems with
continuous variables is far from straightforward. Until now, Schmidt
decomposition in the continuous case required solving the
corresponding integral equations
\cite{schmidtcont,eberly1,eberly2,eberly3,eberly4}. They had to be
discretized, losing the continuous dependence of the initial state.
Here we propose a method to perform the Schmidt decomposition for
this case, to the accuracy desired, keeping the continuous character
of the variables. This method consists of two steps:

1) We decompose the bipartite system wave function, $f(p,q)$, by
using two discrete and complete sets of orthogonal functions,
$\{O^{(1)}_n(p)\}$, $\{O^{(2)}_n(q)\}$, of $L^2$, in the form:

\begin{equation}
f(p,q)=\sum_{m,n}C_{mn}O^{(1)}_m(p)O^{(2)}_n(q)\label{eq11}
\end{equation}

The purpose of this step is to transform the continuous problem
into a discrete one (a necessary step for the numerical
computation), while preserving the continuous dependence of
$f(p,q)$.

2) Then we apply the (finite dimensional) Schmidt procedure to
(\ref{eq11}) in order to write the wave function $f(p,q)$ as
diagonal sum of biorthogonal terms:

\begin{equation}
f(p,q)=\sum_n
\sqrt{\lambda_n}\psi^{(1)}_n(p)\psi^{(2)}_n(q)\label{eq12}
\end{equation}

 The orthogonal functions $\psi^{(1)}_n(p)$, $\psi^{(2)}_n(q)$ -the modes- will be
some particular linear combinations of $O^{(1)}_n(p)$,
$O^{(2)}_n(q)$, respectively. Notice that we are using the Schmidt
procedure for discrete systems to obtain the decomposition for the
continuous case. This is much more tractable, as it implies
diagonalizing matrices instead of solving integral equations.

The rationale for this procedure is the expectation that only a
few $O_n$ will suffice: A handful of appropriate orthogonal
functions will approximate $f(p,q)$ to the desired accuracy. We
finish by pointing out some properties of this method, namely

a) We obtain complete analytic characterization of the modes
$\psi^{(1)}_n(p)$, $\psi^{(2)}_n(q)$ to the desired precision. Our
method surpasses the standard numerical procedures in that keeps
the continuous features present in $f(p,q)$.

b) We remark the portability of the attained modes
$\psi^{(1)}_n(p)$, $\psi^{(2)}_n(q)$ that are ready for later
uses.

c) For the physical system analyzed in this paper (biphoton), we
found that with $26\times 26$ $C_{mn}$ matrices the error was of
around $2\%$, that is, the number of $O_n$ functions required is
small. For other systems studied that we do not include here, the
convergence was even better: For the case of two electrons which
interact electrostatically, the obtained error with $12\times 12$
matrices was of $0.7\%$ (Schmidt number $K=2.4$).

In this paper we begin with a detailed exposition of our method in
Sect. \ref{s2}. Then, in Sect. \ref{s4} we apply it to a relevant
case: the entanglement of two photons created by parametric
down-conversion. We compare our results (leading to well known,
continuous functions) with those computed by numerical methods
\cite{eberly1} that produce sets of points: discrete functions.
Both methods agree remarkably well. Finally, in Sect. \ref{delta}
we decompose the Dirac delta, a case of physical and mathematical
interest.

\section{Schmidt decomposition with discrete sets of orthogonal
functions\label{s2}}

We consider a bipartite quantum system formed by two subsystems
$S_1$ and $S_2$. Some examples are two photons entangled by
parametric down-conversion, a photon emitted by an excited atom
and as a result entangled with it or two charged particles which
interact electrically. This system is described by the vector
state

\begin{eqnarray}
|\psi\rangle=\int dp dq f(p,q)
a^{\dag}_{(1)}(p)a^{\dag}_{(2)}(q)|0,0\rangle\\
(||f(p,q)||^2\equiv\int dp dq |f(p,q)|^2<\infty)\nonumber
\end{eqnarray}

 where
$a^{\dag}_{(1)}(p)$, $a^{\dag}_{(2)}(q)$ are the creation
operators of a particle associated to the subsystems $S_1$ and
$S_2$. $p$ and $q$ are continuous variables associated to $S_1$
and $S_2$ respectively, which can represent momenta, energies,
frequencies, or the like. In general, the analysis is made in an
ad hoc kinematical situation in which $p$ and $q$ turn out to be
one-dimensional variables, $p\in (a_1,b_1)$, $q\in (a_2,b_2)$. In
the following we assume this is the case. In addition, there can
be discrete variables (like the spin) to be treated with the
Schmidt method, that we do not include here to avoid unwieldy
notation.

Our method works as follows:

We consider two denumerable, complete sets of orthogonal $L^2$
functions $\{O^{(1)}_n(p)\}$, $\{O^{(2)}_n(q)\}$
$n=0,1,...,\infty$, each one associated to each particular
subsystem $S_\alpha$ ($\alpha=1,2$). These functions obey

\begin{eqnarray}
\int_{a_\alpha}^{b_\alpha}dk
O^{(\alpha)*}_m(k)O^{(\alpha)}_n(k)=\delta_{mn}\label{eq22}\\
\sum_n
O^{(\alpha)*}_n(k)O^{(\alpha)}_n(k')=\delta(k-k')\label{eq23}
\end{eqnarray}

1) Our first step is to expand the wave function $f(p,q)$ as a
linear combination of the $O^{(\alpha)}_n$, translating the
continuous problem into a discretized one. Thus we work with the
discrete coefficients of the linear combination, though the
continuous character of the state is preserved in the $k$
dependence of the $O^{(\alpha)}_n$ functions. The expansion reads:

\begin{equation}
f(p,q)=\sum_{m,n=0}^{\infty}C_{mn}O^{(1)}_m(p)O^{(2)}_n(q)\label{eq24}
\end{equation}

where the coefficients $C_{mn}$ are given by

\begin{equation}
C_{mn}=\int_{a_1}^{b_1} dp O^{(1)*}_m(p)\int_{a_2}^{b_2} dq
O^{(2)*}_n(q) f(p,q)\label{eq25}
\end{equation}

2) Our second step is to apply the Schmidt decomposition to the
discretized bipartite state (\ref{eq24}), as is usually done for
finite dimension Hilbert spaces (diagonalizing matrices, instead
of solving integral equations). In order to do this, it is
necessary to truncate the expansion (\ref{eq24}), something that
is possible to a certain accuracy due to the fact that $\int dp dq
|f(p,q)|^2<\infty$ ($f(p,q)$ is in principle normalizable), and
the expansion is in orthogonal functions, so the coefficients
$C_{mn}$ go to 0 with increasing $m,n$ (see below).

We truncate the series (\ref{eq24}) at $m=m_0$, $n=n_0$, with
$m_0\leq n_0$, without loss of generality. The Schmidt procedure
leads to (\ref{eq12}), where

\begin{eqnarray}
\psi^{(1)}_i(p)&=&\sum_{m=0}^{m_0} V_{im}O^{(1)}_m(p)\label{eq26}\\
\psi^{(2)}_i(q)&=&\frac{1}{\sqrt{\lambda_i}}\sum_{m=0}^{m_0}\sum_{n=0}^{n_0}
V^*_{im}C_{mn}O^{(2)}_n(q)\label{eq27}\\
i&=&0,...,m_0\nonumber
\end{eqnarray}

Here the matrix $V$ is the (transposed) eigenvectors matrix of
$M_{ij}=M^*_{ji}\equiv \sum_{n=0}^{n_0} C_{in}C^*_{jn}$:

\begin{equation}
\sum_{m=0}^{m_0}M_{im}V_{jm}=\lambda_jV_{ji}\label{eq28}
\end{equation}

and $\{\lambda_i\}_{i=0,...,m_0}$ are the eigenvalues of $M$.

There are two sources of error in this procedure:

a) Truncation error: This is the largest source of error in our
method. Inescapably, the series (\ref{eq24}) must end at some
finite $m$, $n$ when attempting to obtain some specific result.
This step is possible to a certain accuracy because the function
$f(p,q)$ is square-integrable and we are expanding it into
orthogonal functions, so
$\sum_{m=0}^{\infty}\sum_{n=0}^{\infty}|C_{mn}|^2<\infty$ and thus
$C_{mn}\rightarrow 0$ when $m,n\rightarrow\infty$.

The particular choice of the orthogonal functions $O^{(\alpha)}$
will affect how fast the $C_{mn}$ go to zero. Hence, the election
of these functions for a particular physical problem will be a
delicate task. To reach the same accuracy with different sets
$\{O^{(\alpha)}\}$ it will be necessary in general to consider a
different pair of cut-offs $\{m_0,n_0\}$ for each of the sets.

b) Numerical error: This is a better controlled source. It
includes the error in calculating the coefficients $C_{mn}$ via
(\ref{eq25}) and the one produced when diagonalizing the matrix
$M\equiv CC^{\dag}$.

The suitable quantity to control the convergence for a particular
$f(p,q)$ and a specific set $\{O^{(\alpha)}\}$  is the well known
(squared) distance $d^1_{m_0,n_0}$ between the function $f(p,q)$ and
the Schmidt decomposition obtained with cut-offs $\{m_0,n_0\}$ (mean
square error): \pagebreak

\begin{eqnarray}
d^1_{m_0,n_0}\equiv\nonumber\\\frac{\int_{a_1}^{b_1}
dp\int_{a_2}^{b_2} dq|f(p,q)-\sum_{m=0}^{m_0}
\sqrt{\lambda_m}\psi^{(1)}_m(p)\psi^{(2)}_m(q)|^2}{||f(p,q)||^2}\label{eq29}
\end{eqnarray}

this expression gives the truncation error. It will go to zero
with increasing cut-offs according to the specific
$\{O^{(\alpha)}\}$ chosen.

Another easily computable, less precise way of controlling the
convergence is given by the fact that (with no cut-offs)
$\sum_{m=0}^{\infty}\lambda_m=||f(p,q)||^2$ and thus

\begin{equation}
d^2_{m_0,n_0}\equiv
1-\frac{\sum_{m=0}^{m_0}\lambda_m}{||f(p,q)||^2}\label{eq29bis}
\end{equation}

is other measure of the truncation error, where here $\lambda_m$ is
calculated with cut-offs $\{m_0,n_0\}$. Would we compute the
$\lambda_n$ exactly, then $d^1=d^2$. In practice this can not be
done because our $\lambda_n$ are the eigenvalues of the $m_0\times
m_0$ matrix $M_{ij}$, that depend slightly on $m_0, n_0$. Both
distances behave in a very similar way, as we show in Fig.
\ref{fig1} and Fig. \ref{fig1bis}, though $d^2$ is more easily
computable than $d^1$.

 The election of the two sets of orthogonal
functions for a particular physical problem,
$\{O^{(\alpha)}\}_{\alpha=1,2}$ can be approached from two
different points of view, according to the feature one desires to
emphasize: fundamental or practical.

\subsection{Fundamental point of view\label{ss11}}

The election of the orthogonal functions in a particular problem
can be done according to the specific intervals in which the
variables $p,q$ take values for that case. Typical examples of
discrete sets of orthogonal functions are the orthogonal
polynomials, defined in a variety of intervals. For example, a
possible choice to describe one dimensional momenta
$p\in(-\infty,\infty)$,  are the Hermite polynomials,
$O^{(1)}_n(p)\sim H_n(p)$. The equivalence sign indicates here
that the polynomial must be accompanied by the square root of the
weight function in order to be correctly orthonormalized, and
normalization factors must be included. If, on the other hand, the
variable of interest in a specific problem is bounded from below,
like the energy of a free massless particle $p\in(0,\infty)$ ,
then the election could be Laguerre polynomials, $O^{(1)}_n(p)\sim
L_n(p)$.

The criterion for choosing the orthogonal functions $O^{(\alpha)}$
according to the intervals in which $p$, $q$ are defined has a
fundamental character. For example, the localizability in
configuration space of the Fourier transforms of the modes
(\ref{eq26}), (\ref{eq27}), depends critically on the intervals in
which these modes are defined \cite{iwo}. Only if we choose the
functions $O^{(\alpha)}$ to be defined exactly in the same
intervals as the amplitude $f(p,q)$, may the Fourier transforms of
the modes have the right localization properties. In spite of
that, this point of view may not be the most suitable one, as it
may give slower convergence than the point of view presented
below.

\subsection{Practical point of view\label{ss12}}

In this case, the election is approached with the goal of
improving the convergence. The $O^{(\alpha)}$ are chosen here
according to the functional form of $f(p,q)$. The closer the
lowest modes are to $f$ the lesser the number of them necessary to
obtain the required accuracy. What we are looking for here are
$O^{(\alpha)}$ that maximize $\int_{a_1}^{b_1} dp
O^{(1)*}_m(p)\int_{a_2}^{b_2} dq O^{(2)*}_n(q) f(p,q)$ for low
$m$, $n$.

In some cases this practical point of view will be more useful
than the fundamental one. For example, suppose the amplitude for a
particular problem is of the form
$f(p,q)=g(p,q)e^{-(p/\sigma_p)^2}e^{-(q/\sigma_q)^2}$, with
$g(p,q)$ a slowly varying function of $p$, $q$. In this particular
case it is reasonable to choose the functions $O^{(\alpha)}$ as
Hermite polynomials, because their weight functions are indeed
gaussians. This leads to $O^{(1)}_n(p)\propto H_n(p)e^{-p^2/2}$,
and similarly for $O^{(2)}_n(q)$.

We call this approach practical because the convergence is reached
faster. There is a price: the information relevant to localization
might be lost.

\section{Entanglement of continuous variables in Parametric Down-Conversion\label{s4}}

In this section we consider a realistic case of biphotons already
studied in the literature \cite{eberly1,eberly4}: two photons entangled in
frequency through parametric down-conversion. We apply our method
to this physical system in order to obtain the Schmidt
decomposition and the structure of modes without losing the
analytic character within the target accuracy.

The system under study is a biphoton state generated by parametric
down-conversion (PDC) of an ultrashort pump pulse with type-II phase
matching. The amplitude in this particular case takes the form
\cite{eberly1}

\begin{eqnarray}
f(\omega_o,\omega_e)=\exp[-(\omega_o+\omega_e-2
\bar{\omega})^2/{\sigma^2}]\nonumber\\\times
\mathrm{sinc}\{L[(\omega_o-\bar{\omega})
(k'_o-\bar{k})+(\omega_e-\bar{\omega})(k'_e-\bar{k})]/2\}\label{eq41}
\end{eqnarray}

where $\omega_o,\omega_e\in(0,\infty)$ are the frequencies
associated to the ordinary and extraordinary fields respectively,
$k'_o$ and $k'_e$ are the inverse of group velocities at the
frequency $\bar{\omega}$, $\bar{k}$ is the inverse group velocity
at the pump frequency $2\bar{\omega}$, $L$ is the PDC crystal
length and $\sigma$ is the width of the initial pulse. Typical
values for these parameters are  $(\bar{k}-k'_e)L=0.213ps$,
$(\bar{k}-k'_o)L=0.061ps$, $\bar{\omega}=2700ps^{-1}$, $L=0.8mm$
and $\sigma=35ps^{-1}$.

We perform now the following change of variables

\begin{eqnarray}
p&=&\frac{\omega_o-\bar{\omega}}{\sigma};\;\;\;\;
L_p=(\bar{k}-k'_o)L\sigma
\label{eq43}\\
q&=&\frac{\omega_e-\bar{\omega}}{\sigma};\;\;\;\;
L_q=(\bar{k}-k'_e)L\sigma\label{eq45}
\end{eqnarray}

and thus obtain

\begin{equation}
f(p,q)=e^{-(p+q)^2}\mathrm{sinc}[(L_pp+L_qq)/2]\label{eq46}
\end{equation}

We have applied our method to the function (\ref{eq46}) (once
normalized) according to section \ref{s2}, in the following way:

We choose as orthogonal functions Hermite polynomials, because
their weights are gaussians and a gaussian appears in
(\ref{eq46}). These polynomials were used in \cite{walmsley} for PDC in some particular cases which are exactly solvable. The orthonormal sets we chose,
looking for maximizing the $C_{mn}$ (\ref{eq25}) for the lowest
$m$, $n$, were

\begin{equation}
O^{(\alpha)}_n(k)=(\sqrt{\pi}2^nn!)^{-1/2}H_n(k)e^{-k^2/2}\;\;\;\;\alpha=1,2\label{eq47}
\end{equation}

 This choice of polynomials is suitable for the practical
approach (subsection \ref{ss12}), taking into account that
$\bar{\omega}\gg\sigma$ and thus the interval of definition of
$f(\omega_o,\omega_e)$ can be restricted to a region centered in
$\bar{\omega}$ of width $\sim\sigma$ in $\omega_o,\omega_e$. We
did a careful analysis of this, that for brevity we do not show
here. Notice that our conclusions would not apply in the case
$L_p=L_q$.

We have considered cut-offs $m_0=n_0$ taking values $\{5-25\}$ and
followed the steps of Sect. \ref{s2}. We have computed the
eigenvalues $\lambda_n$ of the Schmidt decomposition (\ref{eq12})
for each pair $\{m_0,n_0\}$. We have also computed the modes
(\ref{eq26}) and (\ref{eq27}).

\begin{figure}
\includegraphics{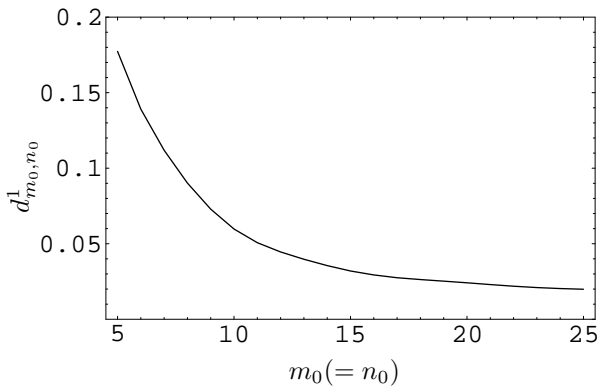}
\caption{$d^1_{m_0,n_0}$ as a function of the cut-offs $\{m_0,n_0\}$.\label{fig1}}
\end{figure}

\begin{figure}
\includegraphics{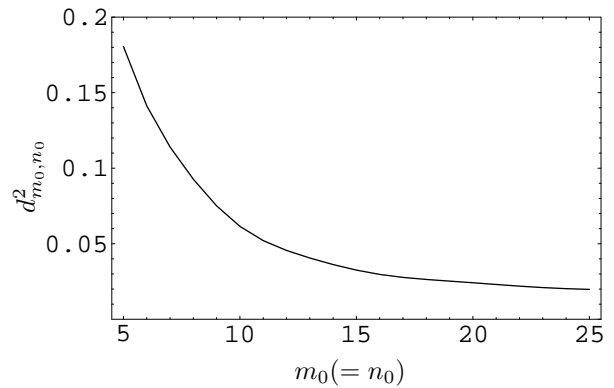}
\caption{$d^2_{m_0,n_0}$ as a function of the cut-offs $\{m_0,n_0\}$.\label{fig1bis}}
\end{figure}

In Fig. \ref{fig1} we plot the distance $d^1_{m_0,n_0}$ (\ref{eq29})
as a function of $m_0=n_0$, to show how fast the convergence is.
With $m_0=n_0=25$ the truncation error is of $2\%$. We also plot in
Fig. \ref{fig1bis} the distance $d^2_{m_0,n_0}$ (\ref{eq29bis}),
which serves as another measure of the convergence, as a function of
$m_0=n_0$. We obtained $d^2_{25,25}=2\%$.

\begin{figure}
\includegraphics{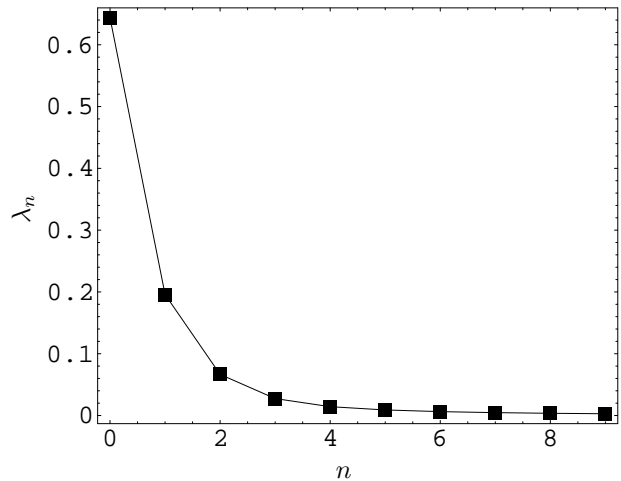}
\caption{Eigenvalues $\lambda_n$ versus index $n$.\label{fig2}}
\end{figure}

\begin{figure}
\includegraphics{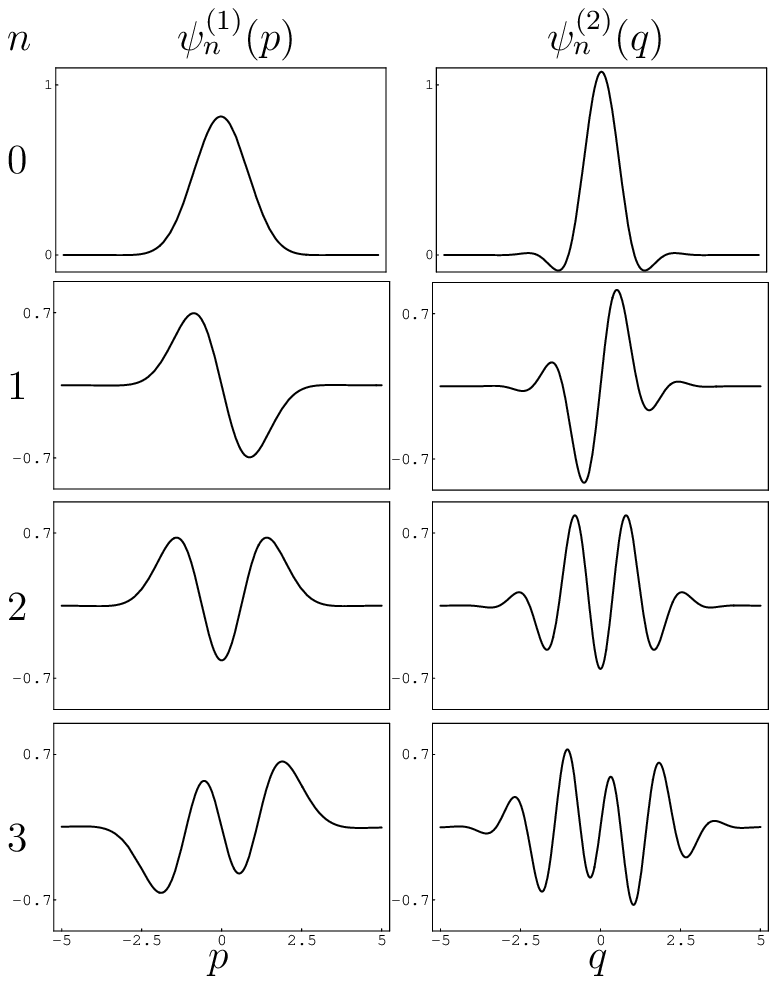}
\caption{Modes $\psi^{(1)}_n(p)$, $\psi^{(2)}_n(q)$ as a function
of $p=\frac{\omega_o-\bar{\omega}}{\sigma}$ and
$q=\frac{\omega_e-\bar{\omega}}{\sigma}$, for
$n=0,1,2,3$.\label{fig3}}
\end{figure}

Regarding now the most precise case considered, $m_0=n_0=25$, we
plot in Fig. \ref{fig2} the eigenvalues $\lambda_n$ for different
values of $n$, observing good agreement with the results existing in
the literature \cite{eberly1}. For this case we also plot in Fig.
\ref{fig3} the modes (\ref{eq26}) and (\ref{eq27}) for $i=0,1,2,3$,
confirming the validity of the method when comparing with
\cite{eberly1}.

The modes are given explicitly by:

\begin{eqnarray}
\psi^{(\alpha)}_m(k)&=&e^{-k^2/2}\sum_{n=0}^{25}(\sqrt{\pi}2^nn!)^{-1/2}
A^{(\alpha)}_{mn}H_n(k)\label{eq49}\\m&=&0,...,25\;\;\;\;\alpha=1,2\nonumber
\end{eqnarray}

where the values of the coefficients $A^{(\alpha)}_{mn}$ are
obtained through (\ref{eq26}) and (\ref{eq27}). The actual
properties of the modes (\ref{eq49}) depend on these values. In
fact, the parity and number of nodes is determined by them, taking
into account that $H_n$ is a polynomial of degree $n$, parity
$(-)^n$ and having $n$ nodes.

 A good approximation to the $\psi^{(1)}_0(p)$ obtained with our
 procedure is

\begin{eqnarray}
\psi^{(1)}_0(p)=e^{-p^2/2}(0.81395-0.14764 p^2+0.00821
p^4)\label{eq410}
\end{eqnarray}

This expression has a deviation (squared distance) of $10^{-5}$
  from the whole mode obtained including terms until $p^{25}$,
   which is the greatest power appearing for $m_0=n_0=25$. On the other hand,
  $d^1_{4,4}-d^1_{25,25}=0.213\gg 10^{-5}$.
      From (\ref{eq410}) it can be seen that in this mode the even
components are greater than the odd ones (these are negligible), so
it is an even state, as shown in Fig. \ref{fig3}.

Another example is the approximation to $\psi^{(2)}_1(q)$

\begin{eqnarray}
\psi^{(2)}_1(q)=e^{-q^2/2}(2.91088 q
        - 3.54070 q^3+1.29062 q^5\nonumber\\
           -0.20402 q^7+ 0.01598 q^9 - 0.00063 q^{11} + 0.00001 q^{13})\nonumber\\\label{eq411}
\end{eqnarray}

This has a deviation (squared distance) of $10^{-4}$ from the whole
mode obtained including terms until $q^{25}$. On the other hand,
$d^1_{13,13}-d^1_{25,25}=0.020\gg 10^{-4}$. More terms are needed in
(\ref{eq411}), because they go to zero more slowly with increasing
powers of $q$. Here the most important components are the odd ones
(the even ones are negligible), leading to an odd parity state, as
shown in Fig. \ref{fig3}.

To show how the convergence of the method depends on the specific
family pairs of orthogonal functions $\{O^{(1)}_n(p)\}$,
$\{O^{(2)}_n(q)\}$ chosen, we consider the cases of Hermite
orthogonal functions depending on a parameter $\beta$ related to the
width of the gaussian, fixed for each family pair:

\begin{equation}
O^{(\alpha)}_n(k)=\frac{\sqrt{\beta}}{\sqrt{(\sqrt{\pi}2^nn!)}}H_n(\beta
k)e^{-(\beta k)^2/2}\;\;\;\;\alpha=1,2\label{eq47}
\end{equation}

We applied our method to the amplitude (\ref{eq46}) with these sets
of orthogonal functions, for $\beta=1.0,0.5,2.0$, and cut-offs
$m_0=n_0=25,20,15,10$. We show in Table \ref{tab1} the values of
$d^2_{m_0,n_0}$ for these specific parameters.

\begin{table}
\begin{center}$d^2_{m_0,n_0}$\\
\begin{tabular}{|c|c|c|c|}
\hline $m_0=n_0$ & $\beta=1.0$ & $\beta=0.5$ & $\beta=2.0$\\
\hline
25 & 0.020 & 0.13 & 0.037\\
\hline
20 & 0.024 & 0.19 & 0.041\\
\hline
15 & 0.032 & 0.27 & 0.050\\
\hline
 10 & 0.062 & 0.38 & 0.064\\
 \hline
\end{tabular}
\end{center}\caption{$d^2_{m_0,n_0}$ for $\beta=1.0,0.5,2.0$ and
$m_0=n_0=25,20,15,10$.}\label{tab1}
\end{table}

Clearly, the convergence is better for the case $\beta=1$, which we
used in the preceding calculations. In case we chose another type of
orthogonal function for (\ref{eq46}) (Laguerre, Legendre,...), the
convergence would have been much worse because of the specific shape
of that amplitude.

\section{Maximum entanglement: The Dirac delta\label{delta}}

Another interesting case is the Dirac delta. Here we have
$f(p,q)=\delta(p-q)$ and we take the same interval $(a,b)$ for $p$
and $q$. We consider complete sets of orthonormal functions
satisfying $O^{(1)}_n(k)=O^{(2)*}_n(k)$. A particular case is when
they are real functions, as for example the typical orthogonal
polynomials (Legendre, Hermite, Laguerre, Chebyshev,...) are. We
must take into account that the Dirac delta is not a function but
a distribution, and indeed is outside $L^2$. However, we can
calculate the $C_{mn}$ and study how much entanglement does this
state have. We obtain straightforwardly $C_{mn}=\delta_{mn}$. This
gives

\begin{equation}
\delta(p-q)=\sum_{n=0}^{\infty}O^{(\alpha)*}_n(p)O^{(\alpha)}_n(q)\label{eq33}
\end{equation}

which is just the resolution of the identity as given in
(\ref{eq23}). The Schmidt decomposition of the Dirac delta is not
unique, because all the weights $\sqrt{\lambda_n}$ are equal to one
(they are degenerate). In fact, the decomposition can be done with
any complete, denumerable set of orthonormal functions, in the form
(\ref{eq23}). This expression can be seen as an infinite
entanglement case, in the sense explained below. The fact that all
the weights are equal to one, makes sense only because we are
considering a distribution, not an $L^2$ state. The sum of the
squares of the weights, which must be equal to the square of the
norm of the function $f(p,q)$, diverges because the Dirac delta is
not square-integrable.

A possible measure of the entanglement of a state $f(p,q)$ in its
Schmidt decomposition (\ref{eq12}) is given by the von Neumann
entropy \cite{nielsen}

\begin{equation}
S=-\sum_{n=0}^{\infty}\lambda_n \log_2\lambda_n\label{eq34}
\end{equation}

This is usually called the entropy of entanglement.

 The state of $L^2$ closer to (\ref{eq33}) is the case of an entangled state with $N$
 diagonal terms with equal $\lambda_n$, when $N$ goes to infinity. To be correctly normalized it verifies
 $\lambda_n=1/N$, $n=0,...,N-1$ and

\begin{equation}
S=-\lim_{N\rightarrow\infty}\sum_{n=0}^{N-1}\frac{1}{N}\log_2\frac{1}{N}=
-\lim_{N\rightarrow\infty}\log_2\frac{1}{N}=\infty\label{eq35}
\end{equation}

This is the maximum entanglement case. This provides an estimate
of the entropy of the Dirac delta (were it in $L^2$).

The origin of our interest in the $\delta$ comes from the fact
that $f(p,q)$ may evolve in time towards a Dirac delta, as it
happens in time dependent perturbation theory of quantum
mechanics. The entanglement in these cases would increase with
time towards its maximum, corresponding to the Dirac delta.

\section{Conclusions}

In this paper we have introduced a method for computing the Schmidt
decomposition of a bipartite state with continuous degrees of
freedom. In the existing literature
\cite{schmidtcont,eberly1,eberly2,eberly3,eberly4} the decomposition
produced sets of points as approximation to the modes. Our method
gives linear combinations of the well known orthogonal functions as
approximation to them. When these functions are chosen properly, a
handful of them is enough to reach the desired accuracy. We
introduce some criteria for the control of convergence and
truncation error. The result of our method for the decomposition of
a biphoton state produced by parametric down-conversion agrees with
the numeric results in the literature \cite{eberly1}. We also touch
on the last stage of evolution of entanglement for determined
systems, analyzing the Dirac delta case.

\section*{ACKNOWLEDGMENTS}
We thank I. Bialynicki-Birula for useful comments and
correspondence. This work was partially supported by the Spanish
Ministerio de Educaci\'on y Ciencia under project BMF 2002-00834.
The work of L. L. was supported by the FPU grant AP2003-0014.


\end{document}